# FusionMAE: large-scale pretrained model to optimize and simplify diagnostic and control of fusion plasma

## Author Information

### Affiliations

1. **Southwestern Institute of Physics, Chengdu, China**
2. **Tsinghua University, Beijing, China**
3. **Nankai University, Tianjin, China**
4. **Zhejiang University, Hangzhou, China**

Zongyu Yang[1], Zhenghao Yang[1&2], Wenjing Tian[1&2], Jiyuan Li[1], Xiang Sun[1], Guohui Zheng[1&3], Songfen Liu[3], Niannian Wu[1&4], Rongpeng Li[4], Zhaohe Xu[1], Bo Li[1], Zhongbing Shi[1], Zhe Gao[2], Wei Chen[1], Xiaoquan Ji[1], Min Xu[1] & Wulyu Zhong[1]

### Contributions

Zongyu Yang contributed by proposing the idea, designing the algorithm, conducting coding and testing, exploring the downstream tasks, discussing the experimental results, and writing the manuscript.

Zhenghao Yang contributed by performing coding and testing and preparing key result figures.

Wenjing Tian contributed by performing coding and testing, reorganizing the experimental results from perspective of large-scale pretrained model, and writing the manuscript.

Jiyuan Li contributed by exploring the downstream task of plasma evolution prediction.

Xiang Sun contributed by exploring the application of FusionMAE on accessing data quality.

Guohui Zheng contributed by exploring the downstream task of plasma equilibrium fitting.

Songfen Liu contributed by exploring the downstream task of plasma equilibrium fitting.

Niannian Wu contributed by exploring the downstream task of plasma evolution prediction.

Rongpeng Li contributed by exploring the downstream task of plasma evolution prediction.

Zhaohe Xu contributed by exploring the downstream task of disruption prediction.

Bo Li contributed by mentor the development of three downstream algorithms.

Zhongbing Shi contributed to diagnostic systems.

Zhe Gao contributed by writing the manuscript and discussing the experimental results.

Wei Chen contributed to the analysis.

Xiaoquan Ji contributed to the experiments.

Min Xu contributed by writing the manuscript and discussing the experimental results.

Wulyu Zhong contributed by writing the manuscript, discussing the experimental results, and managing project progress.

**Corresponding author**

Correspondence to: Zhe Gao(gaozhe@tsinghua.edu.cn), Wulyu Zhong(zhongwl@swip.ac.cn)

## Abstract

In magnetically confined fusion device, the complex, multiscale, and nonlinear dynamics of plasmas necessitate the integration of extensive diagnostic systems to effectively monitor and control plasma behaviour. The complexity and uncertainty arising from these extensive systems and their tangled interrelations has long posed a significant obstacle to the acceleration of fusion energy development. In this work, a large-scale model, fusion masked auto-encoder (FusionMAE) is pre-trained to compress the information from 88 diagnostic signals into a concrete embedding, to provide a unified interface between diagnostic systems and control actuators. Two mechanisms are proposed to ensure a meaningful embedding: compression-reduction and missing-signal reconstruction. Upon completion of pre-training, the model acquires the capability for 'virtual backup diagnosis', enabling the inference of missing diagnostic data with 96.7% reliability. Furthermore, the model demonstrates three emergent capabilities: automatic data analysis, universal control-diagnosis interface, and enhancement of control performance on multiple tasks. This work pioneers large-scale AI model integration in fusion energy, demonstrating how pre-trained embeddings can simplify the system interface, reducing necessary diagnostic systems and optimize operation performance for future fusion reactors.

## Introduction

Fusion energy has long been regarded as a clean and sustainable power source, with the potential to meet the rapidly growing global energy demand while addressing carbon neutrality. Among the various concepts, tokamak stands out as one of the most promising designs for realizing the first commercial fusion reactor, having achieved remarkable milestones in recent years. For instance, the Joint European Torus set a world record by producing 59 megajoules of fusion energy over a 5-second pulse[1]. The Korea Superconducting Tokamak Advanced Research sustained plasma with ion temperatures exceeding 100 million kelvin for 30 seconds[2]. And the EAST and WEST tokamak realized a steady state plasma lasting up to 1,000 and 1320 seconds, respectively[3]. Meanwhile, the International Thermonuclear Experimental Reactor (ITER), one of the world's largest science projects, involving the collaboration of more than 30 nations, is currently under assembly to demonstrate the scientific feasibility of tokamak reactors[4].

Despite the remarkable achievements, the path to a Fusion Power Plant (FPP) is still hindered by several physical and engineering chanllenges[5]. Controlling and optimizing fusion plasma, a non-linear magnetohydrodynamic system, requires the integration of highly intricate, multidisciplinary subsystems[6]. A wide range of diagnostic systems have been developed to measure plasma properties, including current, temperature, density, rotation, and radiation [7]. Additionally, numerous simulation modules have been created to model plasma behavior across different time scales and spatial regions[8]. Furthermore, a significant number of control actuators have been designed to regulate plasma properties[9]. For instance, ITER incorporates roughly 50 distinct plasma diagnostic systems and 70 sets of control actuators[10]. Moreover, the highly complex Integrated Modeling & Analysis Suite (IMAS) system has been established to support experimental data analysis and simulation program execution[11]. In parallel, various control loops are formed through combining different modules, working in coordination to achieve the overall operation of the fusion reactor[12]. Therefore, while the scientific feasibility of fusion energy has been validated, the engineering complexity of fusion power plant remains prohibitively high, as reflected in the extremely long timeline of ITER[13].

In recent years, artificial intelligence (AI) technologies have been widely applied in computer vision, natural language processing(NLP), and diverse scientific fields such as biology, materials science, and meteorology[14-18]. AI has proven its capability in providing end-to-end, streamlined solutions for complex problems. In the fusion domain, AI has also demonstrated numerous successful applications, including predicting major plasma disruptions[19, 20], automating the realization of complex plasma magnetic configurations[21], and achieving kinetic control to avoid plasma instabilities[22]. Although these small-scale models have effectively addressed specific technical challenges in fusion research, their contributions are limited to localized breakthroughs. However, a reduction in the engineering complexity of fusion reactors from a macroscopic perspective would constitute a far more pivotal contribution. If large-scale model techniques can be integrated, their ability to handle diverse tasks in a generalized manner could significantly reduce the operational complexity of fusion devices. Additionally, their capacity to enhance downstream task performance may improve control precision and optimize key plasma parameters, which would greatly accelerate progress in this field.

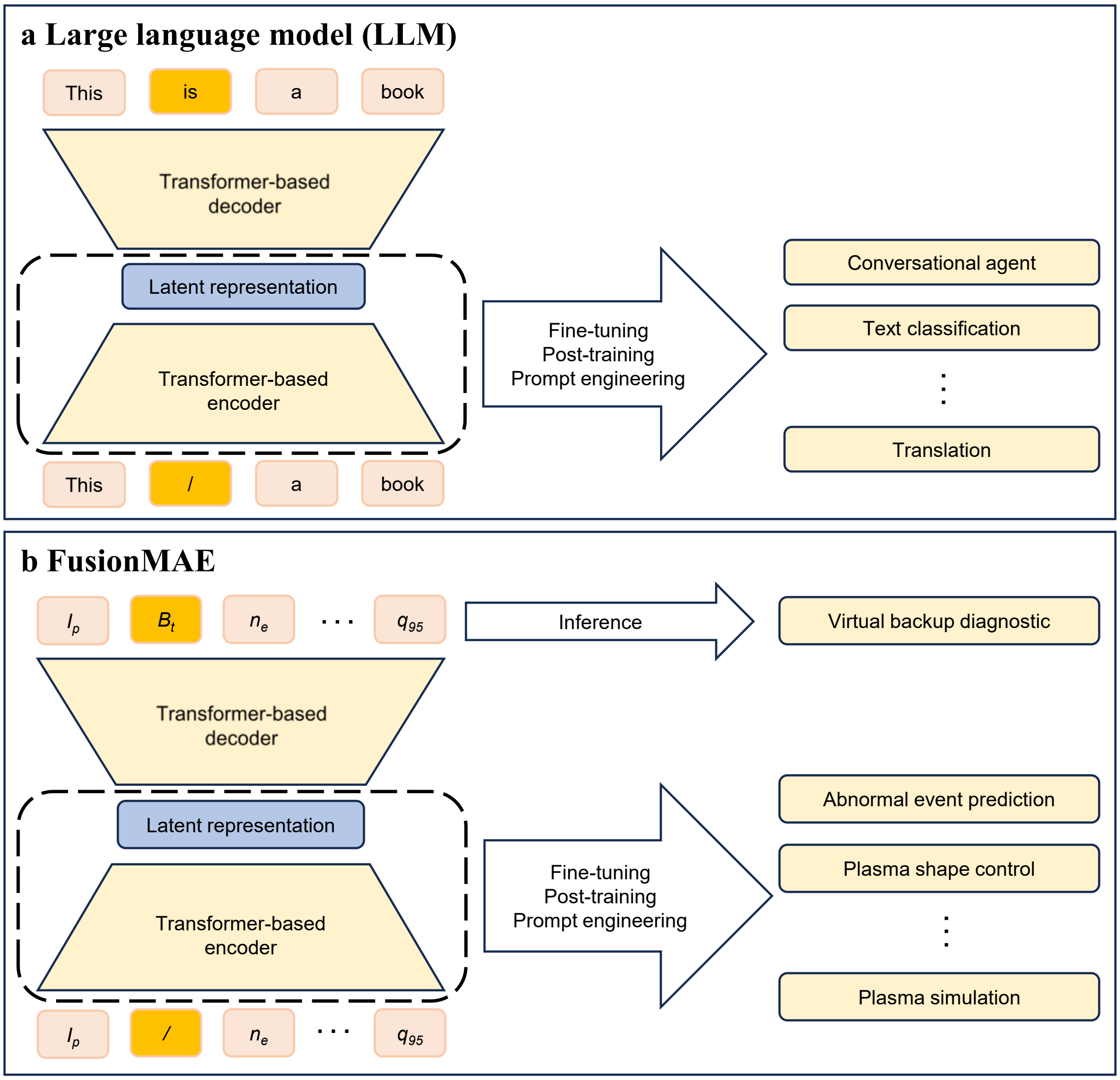

**Fig.1: Comparison between the pre-training and downstream application frameworks of FusionMAE and natural language processing (NLP) large-scale models..**

In this work, we present a pretrained model named the Fusion Masked Auto-Encoder (FusionMAE), which compresses data from multiple diagnostic systems into a unified plasma status embedding, based on a dataset from the largest tokamak currently operated in China, HL-3[23, 24]. This embedding allows a Transformer [25] decoder to restore all the raw diagnostic data and reconstruct possible missing data from it, showing a successfully integration of various diagnostic signals. Analogous to word embeddings in NLP[26], the plasma state embedding naturally clusters discharges according to similarities in operational scenarios. FusionMAE exhibits several compelling emergent capabilities. Firstly, FusionMAE implements automated data analysis by setting the secondary signals to analysis as missing signals and reconstructting them. Secondly, the plasma status embedding can support multiple tokamak operation tasks, including disruption prediction(DPR), equilibrium reconstruction (EFIT-NN) and plasma evolution prediction (PPR) [27-29], showing the possibility to unify the interface between diagnostic systems and control actuators. Surprisingly, using embedding as input even yields superior performance for all tasks compared to using raw data. Thirdly, FusionMAE enables the downstream tasks to perform steadily with several missing diagnostics, improving the robustness of future fusion reactors. It might also enable fusion reactors to reduce the number of necessary diagnostic systems. Figure 1 compares the pre-training and downstream application frameworks of FusionMAE and NLP large-scale models. It demonstrates that FusionMAE is positioned as a domain-specific foundational model designed for fusion experimental data, holding significant potential to evolve into a general-purpose agent tailored to the fusion research.

## Results

### FusionMAE: a Transformer-based model pretrained by self-supervised learning task

FusionMAE is developed to compress 88 signals from 12 systems of HL-3, as shown in Figure 2a, into a unified embedding. The input consists of plasma current ($I_p$), plasma shape and position ($R$, $Z$, $a$, $\kappa$, $\delta_u$, $\delta_l$), toroidal field ($B_t$), poloidal field ($B_p$), loop voltage ($V_{loop}$), electron density ($n_e$), electron temperature ($T_e$), stored energy ($W_e$), radiation power ($P_{rad}$), soft-X radiation ($SX$), impurity radiation($A_{imp}$), D-α spectrum ($D_\alpha$), amplitude of magnetohydrodanamic instability modes ($A_{MHD,\,1}$, $A_{MHD,\,2}$), normalized beta ($\beta_N$), safety factor($q_{95}$), inertial inductive($l_i$), poloidal field coil current ($I_{PF}$) and auxiliary heating power ($P_{NBI}$, $P_{EC}$, $P_{LH}$). Detailed descriptions of these signals are provided in the Methods section. These systems span a broad spectrum of physical parameters, and can be considered as a representative subset of the systems required by future fusion reactors like ITER[10].

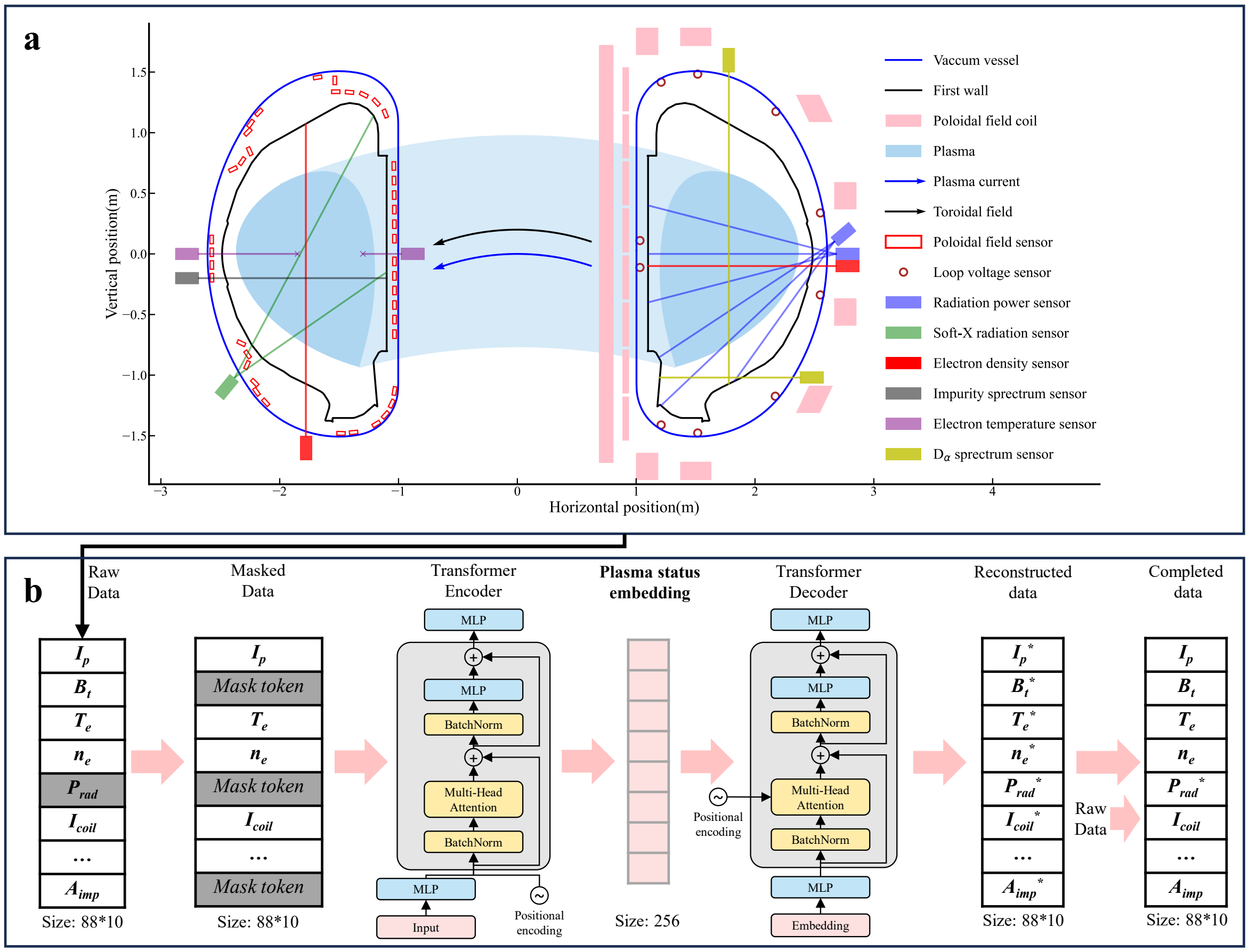

**Fig.2: Overall architecture of FusionMAE.**

**a** Diagram of the diagnostic systems employed in this study. The auxiliary heating systems are omitted due to its complexity, while the details can be found in [30]. Additionally, some secondary parameters, like $A_{MHD}$, $W_e$, $\beta_N$, $q_{95}$ and $l_i$ are computed from a set of primary data using dedicated programs, as described in [31, 32].

**b** Workflow and neural network structure of FusionMAE, which compresses data from multiple diagnostic systems into a unified plasma status embedding.

Fig.2b illustrates the workflow and neural network architecture of FusionMAE. It takes a 10ms time window of the 88 signals sampled at 1kHz as input, forming an 88×10 array. Firstly, each channel of the array is projected into a 64-dimensional space using a multilayer perceptron (MLP), and a positional encoding vector[33] is added to encode the channel index. Secondly, the resulting 88×64 array is then fed to an encoder composed of several Transformer blocks, each consisting of multi-head self-attention layers[34], MLPs, batch normalization layers[35], and residual connections[36]. The encoder take the main responsibility to learn underlying patterns of plasma status. The output of the Transformer encoder is further embedded by an MLP layer into a vector of length 256, which is the expected plasma status embedding. The decoder has a mirrored structure, where several Transformer blocks reconstruct the original data from the embedded vector. The only difference is that the positional embedding is not added, and instead serves as the query vector in the final multi-head attention layer. The entire encoder-decoder structure is trained to minimize the mean squared error between the output and input signals.

## Unified plasma status embedding

FusionMAE extracts a meaningful embedding through two key mechanisms. Firstly, the embedded vector is designed to be smaller than the input array to force effective compression. To preserve information, the model must capture the intrinsic low-dimensional manifold of plasma dynamics. As shown in Fig.3a, a gray radar plot visualizes the Pearson correlation coefficients (PCCs) between the reconstructed and original signals across all 88 channels. The averaged similarity for all channels is 98.6% , indicating that the embedding retains nearly all information contained in the raw diagnostic data, while providing a more compact and structured representation.

However, when the model is solely trained to compress and restore data, it tends to memorize signals on a channel-by-channel basis, and thereby struggles to capture the interconnections between channels. To address this limitation, a more challenging task has been devised. In each training step, a random 25% of the input channels are masked, forcing the model to reconstruct the channels based on its understanding of inter-channel relations. FusionMAE successfully accomplished this task, achieving a PCC of 96.7% for the masked channels, which is presented in Fig.3a by blue radar plot.

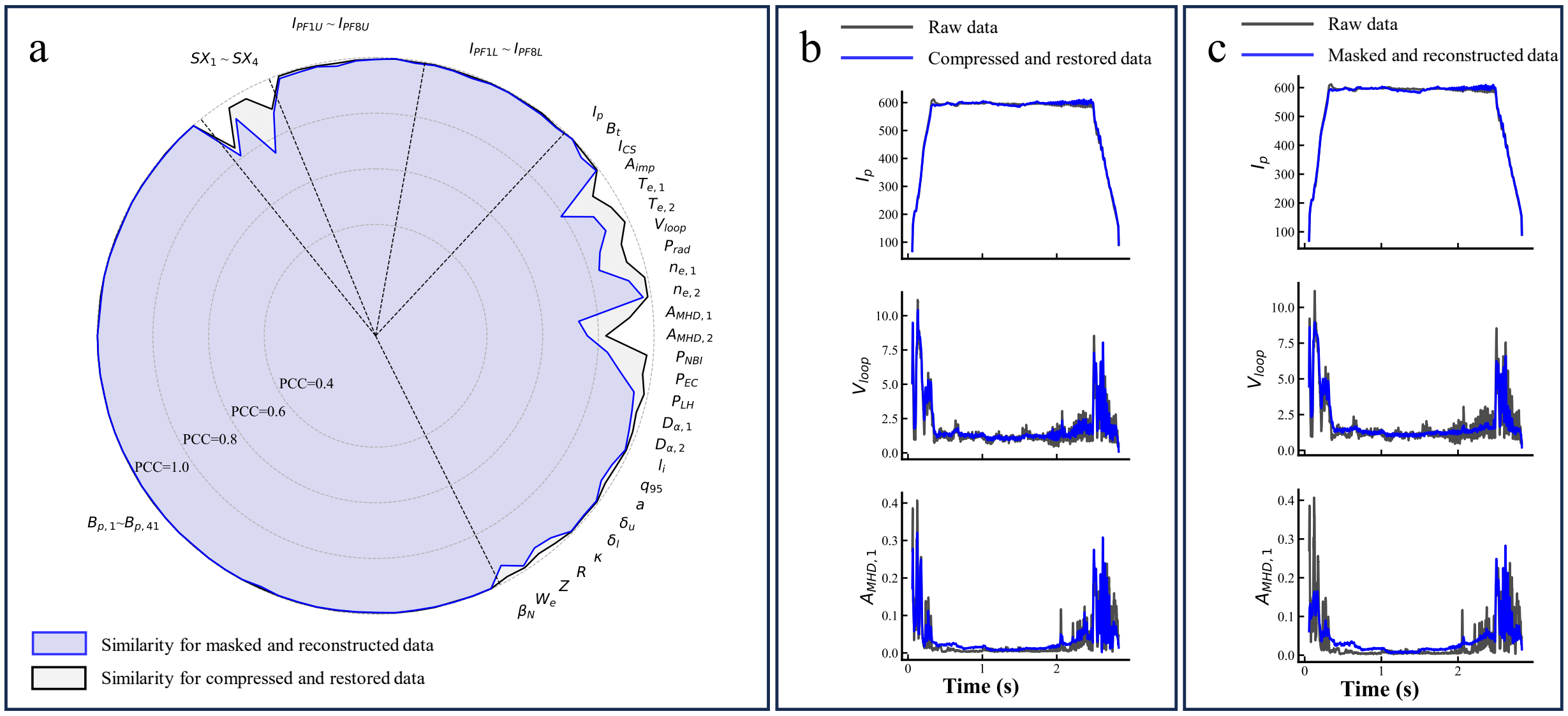

**Fig.3: Evaluation of FusionMAE's performance in restoring compressed raw signals and reconstructing missing signals.**

**a** Radar plot illustrating the mean similarity with raw signals of the compressed-restored signals across 88 channels, as well as the similarity of masked-reconstructed signals. The latter can be seen as the indicator of FusionMAE's reliability in reconstructing missing signals.

**b** Comparison between compressed-restored signals and raw signals for $I_p$, $V_{loop}$ and $A_{MHD,1}$, representing channels with high, medium, and low similarity in (a), respectively.

**c** Comparison between masked-reconstructed signals and raw signals for the same channels as in (b).

The plasma state embedding extracted by FusionMAE plays a role analogous to word embeddings in NLP. Represented by a 256-length vector, it encapsulates the intrinsic physical state of the plasma. Each diagnostic system provides a partial observation of this state, which is transformed into output either through the decoder in digital space, or by physical diagnostic hardware in real experiments. Notably, this state should remain unchanged regardless of the number of diagnostic signals used, while more signals only refine the estimation, and fewer signals result in a coarser approximation. This is also demonstrated in Fig.4a, when the diagnostic signals are randomly reduced by 5% and 20%, the extracted embeddings will only introduce limited oscillation around the black line.

Analogous to word embeddings, the plasma embeddings of physically similar discharges are expected to cluster in latent space. This structure is confirmed via Uniform Manifold Approximation and Projection (UMAP) projection of embeddings from the test set. Notably, despite uniform treatment of all 88 input channels during training, FusionMAE autonomously identifies and organizes embeddings along key plasma parameters ($B_t$, $I_p$, $n_e$ and $\beta_N$), indicating a nontrivial, learned understanding of plasma physics.

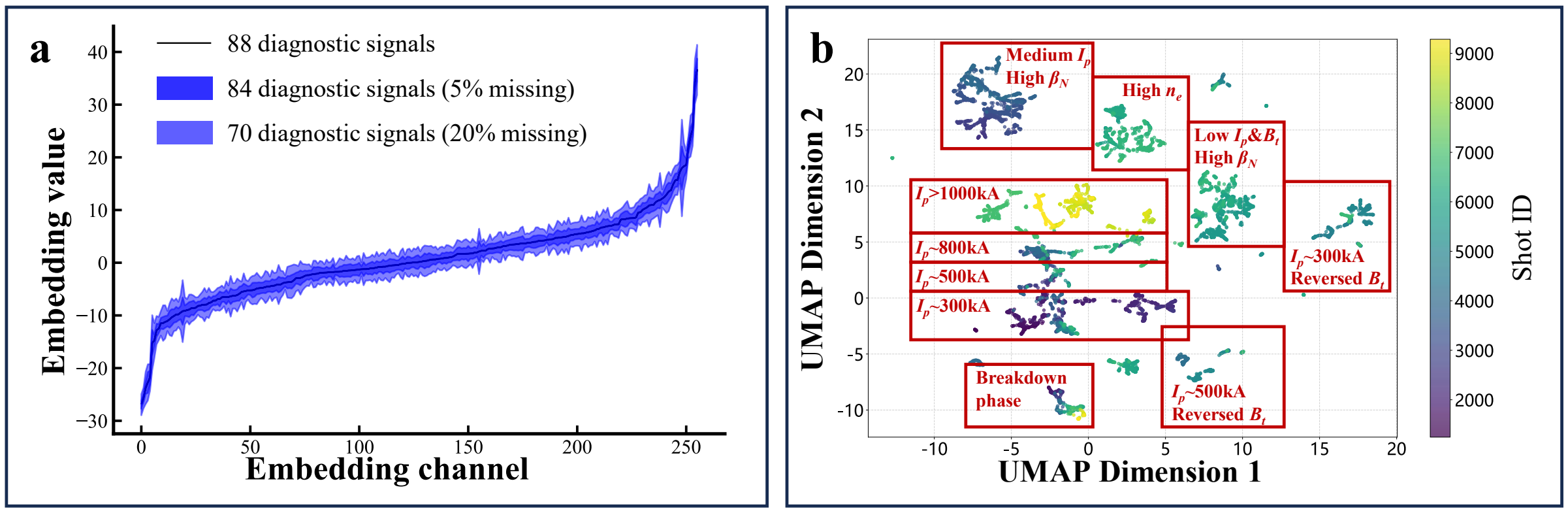

**Fig.4: Demonstration of a plasma status embedding and its similar characteristic as word embedding.**

**a** Plasma state embedding generated by FusionMAE during an HL-3 discharge. Blue bands indicate the uncertainty induced by randomly masking 5% and 20% of the input signals, showing that the embedding remains stable under partial input dropout. Embedding dimensions are sorted by value extracted from complete 88 signals for clarity.

**b** UMAP projection of embeddings from all 100 ms time slices in the test set, revealing clustering aligned with key physical parameters, $B_t$, $I_p$, $n_e$ and $\beta_N$

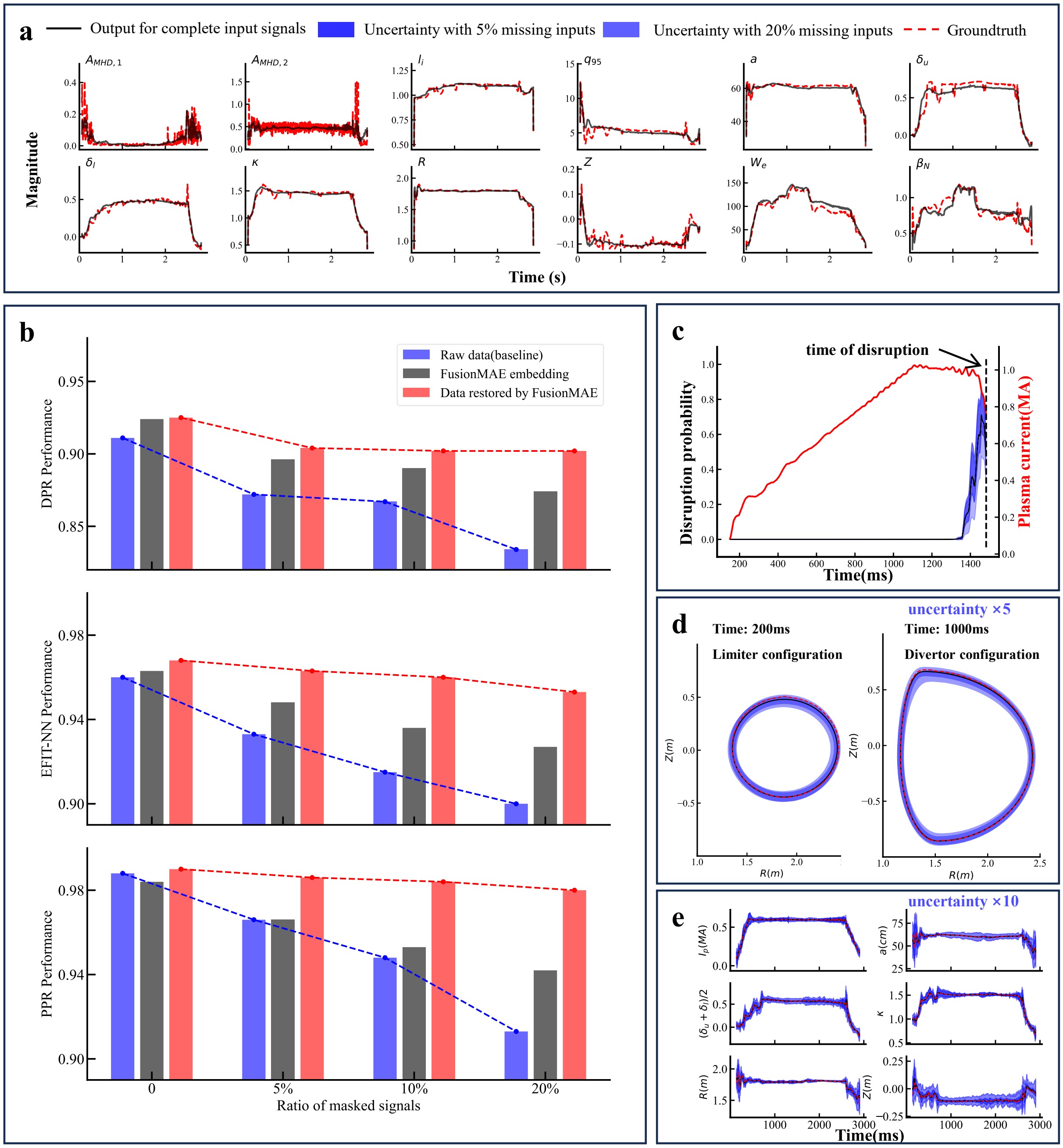

## Fig.5: Emergent capabilities of FusionMAE after pretraining.

**a** Demonstration of using FusionMAE to automatically analyse secondary data. Black lines are reconstructed secondary signals when all of them are masked in the input. Red dashed lines are groundtruth given by traditional analysis tools such as EFIT[37].

**b** The performance of DPR, EFIT-NN and PPR based on models utilizing the plasma status embedding and data completed by FusionMAE, compared with ordinary models that are developed on the raw data.

**c** Output of DPR model during an HL-3 discharge using the plasma state embedding. Blue bands indicate uncertainties introduced by randomly masking 5% and 20% of input signals, demonstrating robustness to missing and invalid diagnostics.

**d** Output of EFIT-NN under the same conditions.

**e** Output of PPR under the same conditions.

## Emerging capability: automatically data analysis for secondary data

While prior analyses have primarily emphasized FusionMAE's pretraining performance, its true value a s a large-scale pretrained model lies in its emergent capabilities. These phenomena, empirically observed, are systematically characterized in this and the following two subsections.

One such capability is signal reconstruction, which can be reframed as a form of automated data analysis. Among the 88 input channels, some are secondary data generated by data analysis programs. For example, the plasma shape parameters ($R$, $Z$, $\kappa$, ...) are calculated by EFIT code based on the poloidal magnetic field and coil currents. An extreme case was constructed where all secondary data were masked and the result is shown in Fig.5a. FusionMAE successfully reconstructed these masked secondary parameters with high fidelity, demonstrating its capacity to infer underlying physical relationships and approximate the functionality of traditional analysis tools

## Emerging capability: all-purpose vector for multiple downstream tasks

Since the plasma status embedding has integrated all the information from the 88 input signals, it is natural to be considered as an 'all-purpose vector' to support various tasks related to tokamak operation.

To demonstrate this capability, three downstream tasks are selected. The first task, disruption prediction (DPR), involves identifying disruption precursors hidden in plasma parameters and exemplifies pattern recognition and anomaly detection[38-41]. The second task, EFIT-NN, is to calculate the plasma shape based on the poloidal magnetic field and coil currents, representing a data analysis challenge[42]. Instead of relying solely on FusionMAE's inherent data analysis capabilities as in Fig.5a, the shape parameters in the 88*10 input array are masked, and an individual neural network is trained to compute them based on the plasma status embedding. The third task, plasma evolution prediction (PPR), is to forecast the future plasma parameters based on the previous parameters and planned coil currents, illustrating predictive simulation[43, 44]. All tasks adhere to the same problem definitions and baseline solutions as described in [27-29], except for certain modifications that conflict with FusionMAE. Performance on DPR is evaluated using the Area Under the Receiver Operating Characteristic Curve (AUC)[45], while EFIT-NN and PPR are measured by the PCC between the predicted outputs and the ground truth[46]. Both AUC and PCC are bounded by 1, with values closer to 1 indicating better performance.

The result of comparison experiments are shown in Fig.5b by the first 3 bars in each subfigure. In these experiments, the performances are tested for three algorithm versions using different inputs, i.e., raw data, plasma status embedding, and data with missing signals completed by FusionMAE. Impressively, both the versions using embedding and completed data achieve similar or even higher performance compared to the version using raw data. These results reveal that FusionMAE provide a precise, self-consistent representation of the plasma and benefits the downstream tasks. The superior performance of the third version over the second one likely stems from some information loss during the generation of the plasma status embedding. By combining the raw data with the reconstructed signals, the third version compensates for this loss and achieves the highest performance.

## Emerging capability: robust control against missing input signals

FusionMAE provides another advantage by enabling robust tokamak operation even when some diagnostic systems are unavailable[47]. On one hand, missing signals can be replaced by mask tokens and subsequently reconstructed by FusionMAE. On the other hand, the embedded vector, which integrates information from multiple diagnostic and control systems, remains stable when part of the input signals are missing.

Fig.5c, 5d, and 5e present the outputs of the DPR, EFIT-NN, and PPR tasks, respectively, during an HL-3 discharge, with the shaded area representing the uncertainty when 5% and 20% of the input is masked. All outputs remain relatively stable, exhibiting only minor oscillations when some input signals are masked. Fig.5b further reports the statistical performance of the DPR, EFIT-NN, and PPR tasks with 5%, 10%, and 20% of the input masked. As expected, the performance of the baseline algorithm degrades rapidly as more signals are masked, whereas the degradation for algorithms supported by FusionMAE is considerably slower. Consequently, FusionMAE establishes a robust operational framework for tokamak reactors while demonstrating potential to minimize diagnostic system dependencies, which is a critical advancement toward FPP.

## Discussion

This study presents FusionMAE, a large-scale pretrained model based on the masked autoencoder (MAE) architecture[48] that captures the intrinsic correlations among diverse plasma and operational parameters in tokamaks, thereby integrating multi-source data into a unified plasma state embedding. For fusion reactor operation, this integrated embedding shows significant potential for enhancing both system integration and reliability.

Regarding system integration, the plasma state vector generated by FusionMAE synthesizes information from multiple diagnostic systems, serving as a universal representation. This vector not only offers robust support for a variety of downstream computational tasks but also exhibits superior feature extraction capabilities that systematically enhance the performance of all downstream modules. Such a unified representation could revolutionize tokamak operational architecture by unifying the interface between diagnostic and control and benefitting all downstream control tasks.

In terms of operational reliability, FusionMAE tackles a critical challenge for future fusion reactors: mitigating operational instability resulting from partial diagnostic failures[49]. The model demonstrates computational resilience when handling incomplete input data, with downstream task outputs showing only bounded fluctuations and quantifiable uncertainty. Notably, FusionMAE achieves high-fidelity data imputation by leveraging its learned understanding of inter-parameter physical relationships, which could be used to reduce the diagnostic demand for future fusion reactors.

From an artificial intelligence perspective, FusionMAE represents a successful application of self-supervised pre-training paradigm [50] to fusion data science. Its emergent properties, including sophisticated feature extraction, cross-task generalization, and intelligent behaviors in data compression and imputation, confirm the applicability of foundational architectures like BERT[51] and GPT[52] to fusion data domains. This finding suggests that fusion-AI integration should transcend conventional small-model approaches (e.g., supervised learning) and embrace cutting-edge techniques from the large-scale model era. Promising directions include applying human feedback reinforcement learning, mixture-of-experts architectures, and chain-of-thought reasoning to develop foundational models for fusion physics[53-56]. The demonstrated success of FusionMAE underscores the potential of adapting state-of-the-art AI methodologies to accelerate progress toward practical fusion energy solutions.

# Methods

## Dataset description

The HL-3 Tokamak, formerly known as HL-2M, is an experimental fusion device developed by the Southwestern Institute of Physics in China. It is designed to enhance the understanding of key physics and engineering principles essential for the advancement of international fusion initiatives such as ITER. HL-3 is capable of sustaining a plasma current between 1.5 and 3 MA while operating under a magnetic field strength of 2.3 T. Its structural characteristics include a major radius of 1.78 m, a minor radius of 0.65 m, and an aspect ratio of approximately 2.8, reflecting its toroidal configuration.

The dataset utilized in this study encompasses 2,445 plasma discharges collected from four HL-3 experimental campaigns, spanning the years 2022 to 2025. The discharges are randomly divided into a development set and a test set, containing 1,950 and 495 shots, respectively. The development set is employed for training and validating the neural network, while the test set is used to assess algorithm performance. As illustrated in Fig.2a, each discharge includes 88 recorded signals that characterize the plasma state. To ensure alignment and streamline the FusionMAE framework, all signals are resampled at 1 kHz. The physical significance, statistical properties, and measurement methods of these signals are detailed in Extended Table 1.

For most channels, the health rate, defined as the proportion of shots in which a given signal channel has recorded valid data, is below 100%. This observation underscores the inherent challenges and complexity of fusion diagnostics, emphasizing that missing diagnostic signals are an unavoidable reality in fusion facilities. To systematically evaluate the performance of various algorithms under different levels of missing data, we constructed three extended datasets based on the original dataset. Specifically, 5%, 10%, and 20% of the signal channels were randomly replaced with Gaussian noise curves, simulating artificial data loss. These extended datasets were subsequently used for comparative experiments, as presented in Fig.4 and 5. The results demonstrate that FusionMAE effectively learns from incomplete datasets, overcoming a key limitation of conventional algorithms, which typically require fully clean and complete data for training and inference.

**Extended Table. 1: The detailed information about the 88 channels used in the development of FusionMAE, including the physical significance, statistical properties and data health rate.**

| Channel name | Channel number | Unit | Physical significance | Statistical properties | | |
|---|---|---|---|---|---|---|
| | | | | Mean value | Standard deviation | Health rate |
| $I_p$ | 1 | MA | Toroidal plasma current, measured by Rogowski coils. | 0.384 | 0.228 | 100% |
| $R$ | 1 | m | Horizontal position of the plasma center in the poloidal cross-section, calculated using the EFIT code. | 1.76 | 0.110 | 90.9% |
| $Z$ | 1 | m | Vertical position of the plasma center in the poloidal cross-section, calculated using the EFIT code. | -0.0608 | 0.113 | 90.9% |
| $a$ | 1 | cm | Minor radius of the plasma in the poloidal cross-section, calculated using the EFIT code. | 58.7 | 7.68 | 90.9% |
| $\kappa$ | 1 | - | Elongation ratio of the plasma in the vertical direction, calculated using the EFIT code. | 1.22 | 0.268 | 90.9% |
| $\delta_u$ | 1 | - | Upper triangularity, defined as the degree to which the upper half of the plasma deviates from a perfectly circular shape towards a triangular form, calculated using the EFIT code. | 0.195 | 0.225 | 90.9% |
| $\delta_l$ | 1 | - | Lower triangularity, defined as the degree to which the lower half of the plasma deviates from a perfectly circular shape towards a triangular form, calculated using the EFIT code. | 0.361 | 0.405 | 90.9% |
| $B_t$ | 1 | T | Toroidal magnetic field strength at the geometric center of the vacuum vessel. | 1.31 | 0.296 | 99.7% |
| $B_p$ | 41 | a.u. | Local poloidal magnetic field strength, measured by 41 pick-up coils surrounding the plasma. | 0.0701 | 0.0777 | 99.8% |
| $V_{loop}$ | 1 | V | Toroidal loop voltage of the plasma. | 2.35 | 3.14 | 100% |
| $n_e$ | 2 | $10^{19}/m^3$ | Line-averaged electron density, measured using Faraday interferometers. | 0.943 | 0.985 | 64.4% |
| $T_e$ | 2 | keV | Electron temperature, measured using electron cyclotron emission (ECE) diagnostics. | 0.428 | 0.454 | 57.3% |
| $W_e$ | 1 | kJ | Plasma stored energy. | 67.7 | 67.8 | 90.9% |
| $P_{rad}$ | 1 | MW | Total thermal radiation power emitted by the plasma. | 0.122 | 0.194 | 63.2% |
| $SX$ | 4 | a.u. | Line-integrated soft X-ray intensity from the plasma. | 0.304 | 0.360 | 78.7% |
| $D_\alpha$ | 2 | a.u. | Intensity of the D-α spectrum. | 0.225 | 0.525 | 74.3% |
| $A_{MHD}$ | 2 | a.u. | Amplitude of magnetohydrodynamic (MHD) instabilities with a toroidal mode number n=1, derived from raw signals recorded by Mirnov probes. | 0.173 | 0.292 | 93.1% |
| $A_{imp}$ | 1 | a.u. | Intensity of the C III impurity spectrum. | 0.141 | 0.270 | 64.7% |
| $\beta_N$ | 1 | - | Normalized beta, representing the normalized plasma pressure. | 0.655 | 0.525 | 90.9% |
| $q_{95}$ | 1 | - | Safety factor at the 95% flux surface. | 5.31 | 2.74 | 90.9% |
| $l_i$ | 1 | - | Internal inductance of the plasma. | 1.03 | 0.068 | 90.9% |
| $I_{PF}$ | 16 | kA | Current in the poloidal field coils of the tokamak. | -0.618 | 1.70 | 99.9% |
| $I_{CS}$ | 1 | kA | Current in the central solenoid coil of the tokamak. | 6.35 | 36.7 | 99.9% |
| $P_{NBI}$ | 1 | kW | Power of neutral beam injection. | 55.1 | 220 | 99.9% |
| $P_{EC}$ | 1 | kW | Power of electron cyclotron resonant heating. | 19.2 | 109 | 98.7% |
| $P_{LH}$ | 1 | kW | Power of lower hybrid resonant heating. | 5.10 | 30.0 | 98.6% |

## Model architecture of FusionMAE

FusionMAE is a Transformer-based masked autoencoder, with its architecture illustrated in Fig.2b. The model design is inspired by the masked autoencoder for vision Transformers proposed in [48]. It processes an input consisting of an 88×10 array, where each row corresponds to one of the 88 diagnostic channels, and each column represents a 10 ms time window of signals sampled at 1 kHz. Within the Transformer framework, each column is treated as a patch, encapsulating plasma information from a single diagnostic channel.

To handle missing and invalid channels, a mask token is introduced in the neural network, replacing unavailable data. This mask token is a randomly initialized trainable vector, which is updated throughout the training process and gradually converges into an optimal representation that encodes missing information. In addition to these inherently missing or invalid channels, 25% of the valid channels are intentionally masked during training. These masked channels serve as the primary targets for reconstruction, enabling FusionMAE to learn to reconstruct missing information effectively, as detailed in the next section.

Each patch is projected into a 64-dimensional space using a multilayer perceptron (MLP), followed by the addition of a positional encoding ($\boldsymbol{PE}$) vector [33] to encode the channel index and inform the encoder about the diagnostic source of each patch. The positional encoding is computed using Equations (1) and (2), where idx represents the index of the input channel, and $\boldsymbol{2i}$ and $\boldsymbol{2i+1}$ refer to the indices of the even and odd elements of the positional encoding vector, respectively.

$$PE(idx, 2i) = \sin(idx/88^{2i/64}) \qquad i = 0,1,2,\ldots,31 \qquad (1)$$

$$PE(idx, 2i+1) = \sin(idx/88^{(2i+1)/64}) \qquad i = 0,1,2,\ldots,31 \qquad (2)$$

The resulting 88×64 array is then passed through an encoder composed of multiple Transformer blocks, each incorporating multi-head self-attention layers [34], MLPs, batch normalization layers [35], and residual connections [36]. This Transformer block architecture is widely adopted in large-scale models such as BERT and GPT, demonstrating its effectiveness in learning intricate dependencies within sequential data.

The encoder primarily captures underlying patterns in plasma behavior. Its output maintains the same shape as the input (88×64) and is subsequently flattened into a 5632-dimensional vector. This vector is further compressed by an MLP into a 256-dimensional embedded representation.

The decoder adopts a structure that closely mirrors the encoder. The 256-dimensional embedded vector is first expanded back into a 5632-dimensional vector, which is then reshaped into an 88×64 array. Subsequently, multiple Transformer blocks process this array to reconstruct the original data.

The key distinction between the encoder and decoder lies in the treatment of positional encoding. Unlike in the encoder, where positional embeddings are directly added to the input patches, the decoder does not incorporate positional embeddings in the same manner. Instead, they serve as query vectors in the final multi-head attention layer, aiding in the reconstruction process.

Finally, an MLP layer projects the 88×64 array into an 88×10 array, representing the reconstructed diagnostic signals across all channels, including those that were originally missing or invalid.

**Extended Table. 2: The detailed information about model structure of FusionMAE**

| Number | Repeat Time | Layer type | Output shape | Layer parameters |
|---|---|---|---|---|
| 1 | 1 | Input layer | 88*10 | |
| 2 | 1 | Multi-layer perceptron | 88*64 | 64 units, dropout rate = 0.2 |
| 3 | 3 | BatchNorm | 88*64 | momentum = 0.99 |
| 4 | | Multi-head attention | 88*64 | 8 heads, key dimention = 64, dropout rate = 0.2 |
| 5 | | BatchNorm | 88*64 | momentum = 0.99 |
| 6 | | Multi-layer perceptron | 88*128 | 128 units, dropout rate = 0.2 |
| 7 | | Multi-layer perceptron | 88*64 | 64 units, dropout rate = 0.2 |
| 8 | 1 | Reshape | 5632 | |
| 9 | 1 | Multi-layer perceptron | 256 | 256 units, dropout rate = 0.2 |
| 10 | 1 | Multi-layer perceptron | 5632 | 5632 units, dropout rate = 0.2 |
| 11 | 1 | Reshape | 88*64 | |
| 12 | 2 | BatchNorm | 88*64 | momentum = 0.99 |
| 13 | | Multi-head attention | 88*64 | 8 heads, key dimention = 64, dropout rate = 0.2 |
| 14 | | BatchNorm | 88*64 | momentum = 0.99 |
| 15 | | Multi-layer perceptron | 88*128 | 128 units, dropout rate = 0.2 |
| 16 | | Multi-layer perceptron | 88*64 | 64 units, dropout rate = 0.2 |
| 17 | 1 | Multi-layer perceptron | 88*10 | 10 units, dropout rate = 0.2 |

## Training of FusionMAE

The FusionMAE model is trained using the AdamW optimizer, aiming to minimize the loss function defined in Equation (3). In this formulation, $X_{recontruct}$ represents the output of FusionMAE. $X_{original}$ denotes the original input data. The term $W$ assigns weights to each input channel, with values set as 0.2, 0, and 1 for valid channels, invalid/missing channels, and intentionally masked channels, respectively.

This weighting strategy ensures that FusionMAE primarily learns to reconstruct the intentionally masked channels, while simultaneously preserving valid information during the compression process. Given the model's architecture, its capability to reconstruct intentionally masked channels naturally extends to the reconstruction of invalid and missing channels, as these are treated equivalently during training.

The detailed hyperparameter settings used for training FusionMAE are summarized in Table 3.

$$loss = \sum_{1}^{88} W * \left|X_{recontruct} - X_{original}\right| \quad (3)$$

## Extended Table. 3: Hyperparameter settings used for training FusionMAE

| Hyper-parameter | Explanation | Value |
|---|---|---|
| Gaussian noise | A Gaussian-distributed noise signal is added to the input data to implement data augmentation during the training phase. | $\mu$=0， $\sigma$=0.003 |
| Optimization method | The function used to map the partial derivatives of the prediction error with respect to the neural network weights to the adjustments made to the neural network weights. | Adam (adaptive moment estimation), with learning rate = $0.001 \times (0.1)^{i/10000}$, in the $i_{th}$ step of training |
| Weight decay | The technique of reducing the absolute values of network weights with a relative ratio after each training step to mitigate overfitting by penalizing excessively large weights. | $0.0003 \times (0.1)^{i/20000}$, in the $i_{th}$ step of training |
| Batch size | The number of samples used to compute the prediction error and update the network weights simultaneously during a training step. | 256 |
| Early stopping | The number of epochs that the training framework will wait for improved performance before halting training and reverting to the network's best weights. | 20 |

## FusionMAE's performance under different mask ratios

While FusionMAE demonstrates effective capabilities to reconstruct missing data, its performance has limits and can still fail when faced with extremely high rates of missing diagnostic data. As shown in Fig.6, at 0% masking (complete diagnostics), near-zero reconstruction loss validates the model's lossless compression capability. Conversely, under 100% masking (complete information loss), the model strategically defaults to output average value on training-set for every channel, which is a mathematically optimal method to minimize the training loss. Intermediate masking ratios reveal two distinct operational regimes. Firstly, below critical masking thresholds (<50%), testing loss progression indicates effective cross-channel correlation exploitation, demonstrating sufficient redundancy in current HL-3 diagnostics. Secondly, when beyond this threshold, non-linear loss escalation exposes fundamental information limits where plasma state reconstruction becomes underdetermined. This phase diagram establishes a quantitative framework for assessing current diagnostic redundancy while providing empirical guidelines for minimum viable diagnostic configurations in future reactors, which is particularly crucial for balancing measurement robustness against system complexity in future fusion reactors.

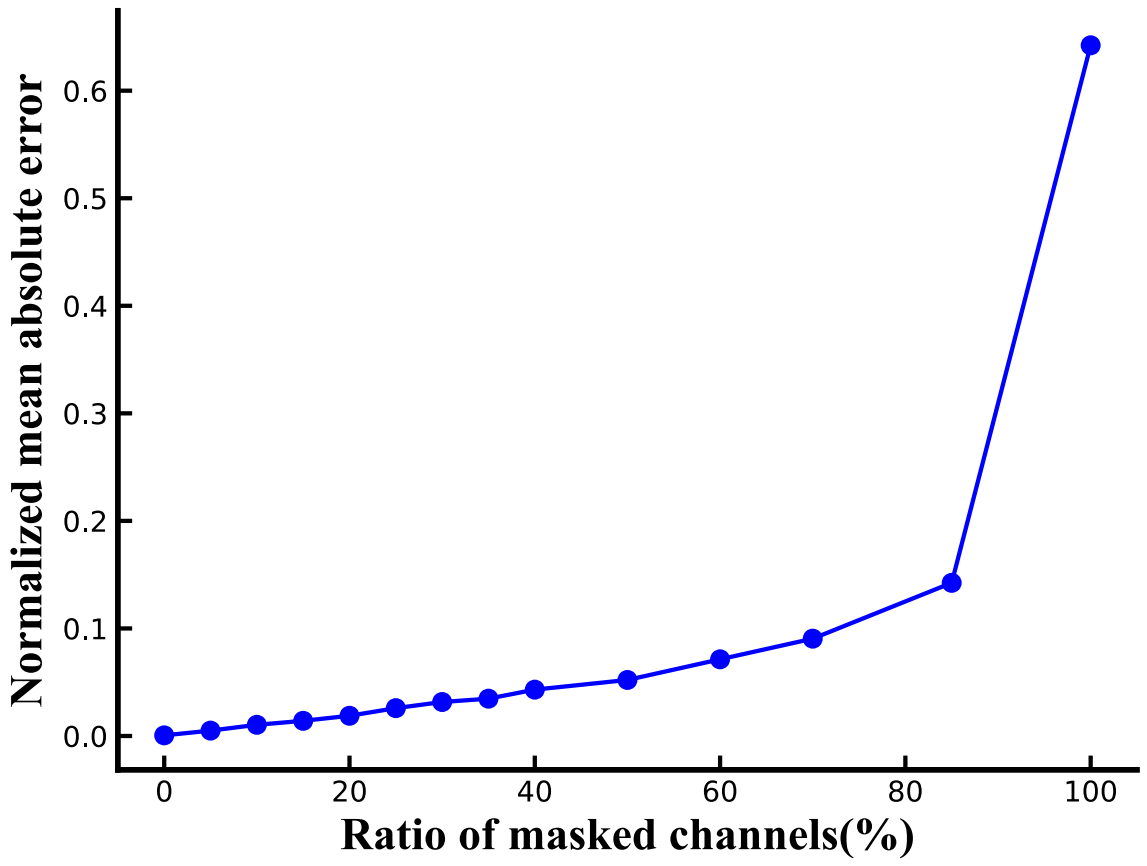

**Extended Fig.6: FusionMAE's testing loss, evaluated by normalized mean absolute error, under different settings of masked ratios during training.**

## Downstream task: disruption prediction

Disruption is an anomalous event that occurs during tokamak discharges, characterized by the sudden loss of plasma confinement. This phenomenon can lead to severe consequences, including intense electromagnetic forces, thermal loads, and the generation of runaway electrons, all of which may cause significant damage to critical components of the device. The primary objective of disruption prediction is to identify precursors of disruptions, such as loss of plasma control and the development of magnetohydrodynamic (MHD) instabilities, thereby enabling the activation of mitigation strategies to minimize potential damage.

The neural network architecture of the disruption prediction algorithms proposed in this study is presented in Fig.7a. The design of algorithm follows a similar paradigm to a prior study conducted on HL-3 [27]. As the compared three version in Fig.5, the baseline version takes time series from 88 diagnostics as input. And in the second version, these signals are replaced by series of 256-dimensional plasma status embedding. In the final version, the input remains the 88 diagnostic signals; however, any missing or invalid signals are replaced by data provided by FusionMAE.

The output of the disruption prediction model is trained to align with the labels defined by Equation (4), where fuzzy means that these data won't be used to update the parameters of neural network, and TTD denotes the time interval between input data and the disruption event. The performance of the algorithms is assessed using the Area Under the Receiver Operating Characteristic Curve (AUC), a widely adopted metric for evaluating disruption prediction models. A detailed definition of AUC can be found in [27].

$$Label = \begin{cases} 1 & TTD < 50ms \\ fuzzy & 50ms < TTD < 300ms \\ 0 & TTD > 300ms \end{cases} \tag{4}$$

## Downstream task: equilibrium fitting

Equilibrium fitting is a fundamental data analysis task in tokamak operation. By numerically solving the Grad-Shafranov equation under the constraints of diagnostic signals—such as the poloidal magnetic field surrounding the plasma—the plasma's flux and current density distribution can be obtained. Based on these distributions, the plasma shape can be determined. The resulting equilibrium solution serves as a critical input for plasma shape and position control, laying the foundation for stable plasma discharge.

In this study, the second downstream task involves training a neural network to serve as a surrogate for the equilibrium fitting program (EFIT). The baseline neural network takes $I_{PF}$, $I_{CS}$, $B_p$ and $I_p$ as inputs and predicts the shape of plasma. Two additional versions of the model employ an embedding vector and reconstructed inputs to perform the same task. The architecture of the surrogate model is illustrated in Fig.7b. The performance of the algorithms is evaluated using the Pearson Correlation Coefficient (PCC), which measures the agreement between the predicted shape parameters and the ground-truth values obtained from EFIT. The implementation details of this downstream task closely follow those of the preceding work in [28].

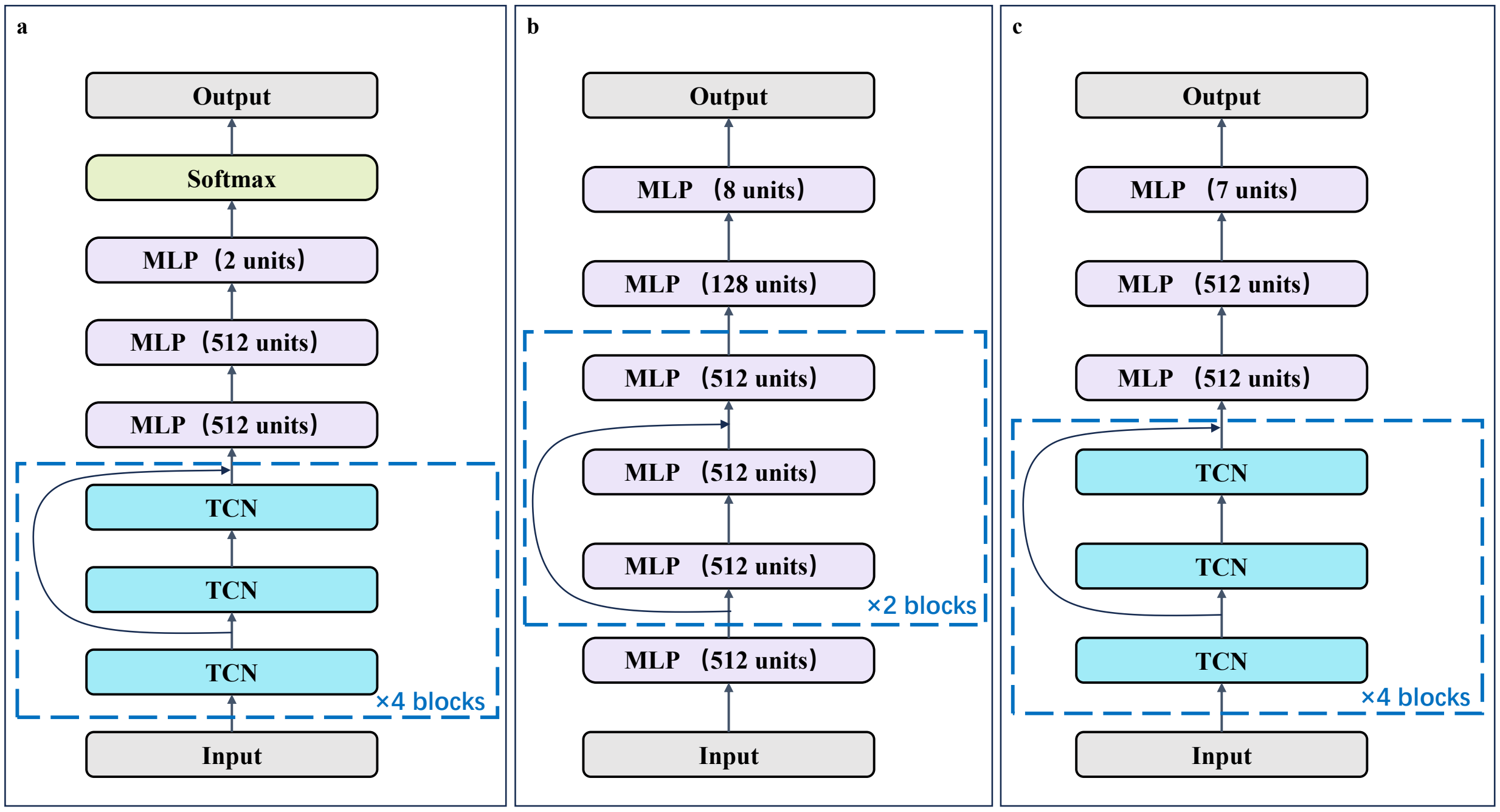

**Extended Fig.7: The neural network structure used in three downstream tasks: (a) disruption prediction, (b) equilibrium fitting surrogate model, (c) plasma parameter prediction. TCN refers to temporal convolutional neural network, whose detailed hyper-parameters keeps the same as in [27, 28]**

## Downstream task: plasma evolution prediction

Plasma evolution prediction serves as the foundation for controller design in tokamak operation. The prediction algorithm can be represented as a function, as described in Equation (5).

$$s(t+1) = F(s(t), a(t)) \qquad (5)$$

where $s(t)$ and $s(t+1)$ denote the present and future plasma parameters, respectively, while a(t) represents the control action. The function $F$ encapsulates the plasma response to various control actions.

Researchers design controllers based on the prediction algorithm to regulate plasma behavior and achieve the desired target parameters. In this research, $s(t)$ and $s(t+1)$ include $I_p$, $R$, $Z$, $a$, $\kappa$, $\delta_u$, $\delta_l$, while the control actions $a(t)$ are represented by $I_{PF}$ and $I_{CS}$. A neural network is trained on historical data to approximate the function $F$ and predict plasma current and shape evolution. Fig.7c illustrates the architecture of the surrogate model. The performance of the algorithms is evaluated using PCC, which quantifies the agreement between the predicted shape parameters and the ground-truth values obtained from experimental results. The implementation details of this downstream task closely follow those of the preceding work in [29].

## Contribution of each diagnostic for plasma status embedding

FusionMAE introduces a novel paradigm for describing plasma status. Rather than relying on an increasing number of diagnostic measurements to provide additional parameters, FusionMAE represents plasma status using a fixed-length vector. As more diagnostic signals are incorporated, the resulting vector becomes increasingly refined, enabling a more precise representation of the plasma state. Consequently, the contribution of each diagnostic signal to defining the plasma status can be quantified using Equation (6):

$$contribution(i) = \frac{1 - PCC^*(i)}{\sum_{j=1}^{88}\left(1 - PCC^*(j)\right)} \qquad (6)$$

where $PCC^*(i)$ denotes the PCC between the embedding vector computed from the complete set of input channels and the embedding vector obtained when masking the $i_{th}$ channel. Fig.8 illustrates the contributions of the 88 diagnostic signals in determining plasma status, providing critical insights into the relative importance of different subsystems for future fusion reactors. For most input channels, this sequential distribution aligns well with physical intuition. The description of plasma state is primarily determined by the most critical physical quantities such as $B_p$, $T_e$, $n_e$ and $B_t$. Secondly, it is directly governed by the dominant terms among external control inputs like $I_{PF}$ and auxiliary heating power. The information about plasma magnetic configuration and current appears somewhat anomalously positioned at the end, but this can be understood as their information being entirely replaceable by $B_p$ and $I_{PF}$. This analysis helps us quantify the contribution of each diagnostic to plasma observation, thereby guiding subsequent optimization of diagnostic systems and the design of fusion reactor diagnostics.

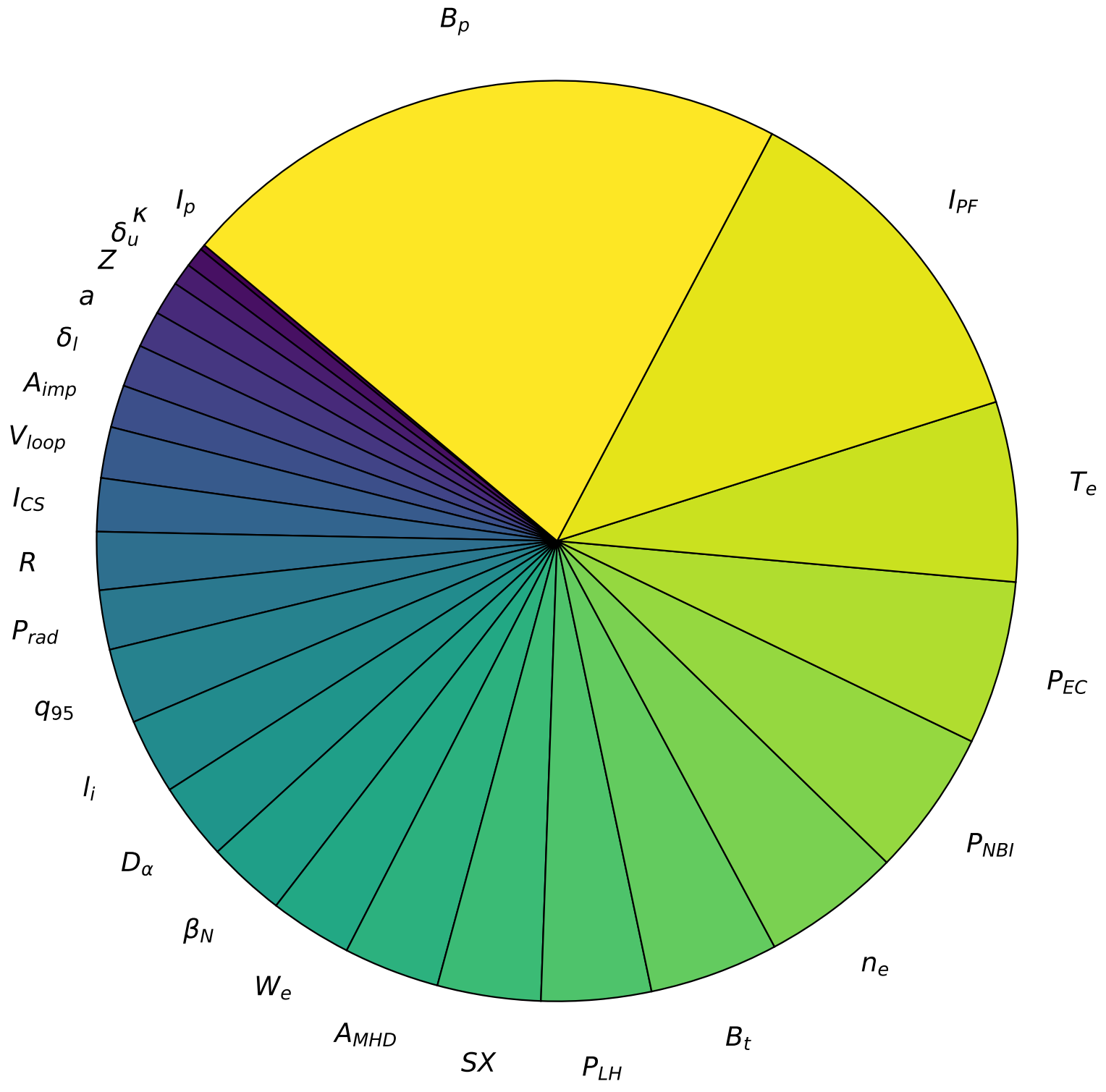

**Extended Fig.8: The contribution of different channels in the determination of plasma status embedding.**

## Data Availability

The data that support the findings of this study are available from the corresponding author upon reasonable request.

## Code Availability

The code that support the findings of this study are available from the corresponding author upon reasonable request.

## Acknowledgements

This work is supported by National MCF R&D program of China under Grant No. 2024YFE03240100 and National Natural Science Foundation of China under Grant No. U21A20440. The authors wish to thank all the members at Southwestern Institute of Physics for their contributions to the collection of dataset and development of algorithm.